\documentclass[11pt,twoside]{article}

\usepackage{asp2006}
\usepackage{epsf}
\usepackage{psfig}
\usepackage{lscape}
\usepackage{graphicx}

\begin{document}
\title{The (Un)Lonely Planet Guide:  Formation and Evolution 
of Planetary Systems from a `Blue Dots' Perspective}   
\author{Michael R. Meyer}   
\affil{Institute for Astronomy, ETH, Z\"urich, Switzerland}    

\begin{abstract} 
In this contribution I summarize some recent successes, and focus on remaining challenges, in understanding the formation and evolution of planetary systems in the context of the Blue Dots initiative.  Because our understanding is incomplete, we cannot yet articulate a 'design reference mission' engineering matrix suitable for an exploration mission where success is defined as obtaining a spectrum of a potentially habitable world around a nearby star.  However, as progress accelerates, we can identify observational programs that would address fundamental scientific questions through hypothesis testing such that the null result is interesting.
\end{abstract}


\section{Introduction}   

Planet formation is a complex process, yet one that can capture the imagination of a hurried public in a busy place (Figure 1).  In our own Solar System, we can identify several possible  outcomes such as the terrestrial planets at orbital radii inside the asteroid belt, the gas giants Jupiter and Saturn, the ice giants Uranus and Neptune, objects in the Edgeworth-Kuiper Belt, and comets.  Without entering into debate concerning ``what is a planet" suffice it to say that there are several flavors of planet formation. And this diversity is expanded greatly when we consider the range of exoplanets discovered around other stars reported elsewhere in this volume.  These planets are thought to have formed in circumstellar disks of gas and dust which initially surrounded the young Sun and indeed most sun-like stars \footnote{Even the major satellites of the giant planets likely formed from circumplanetary disks.}.  

\begin{figure}[!ht]
\plotone{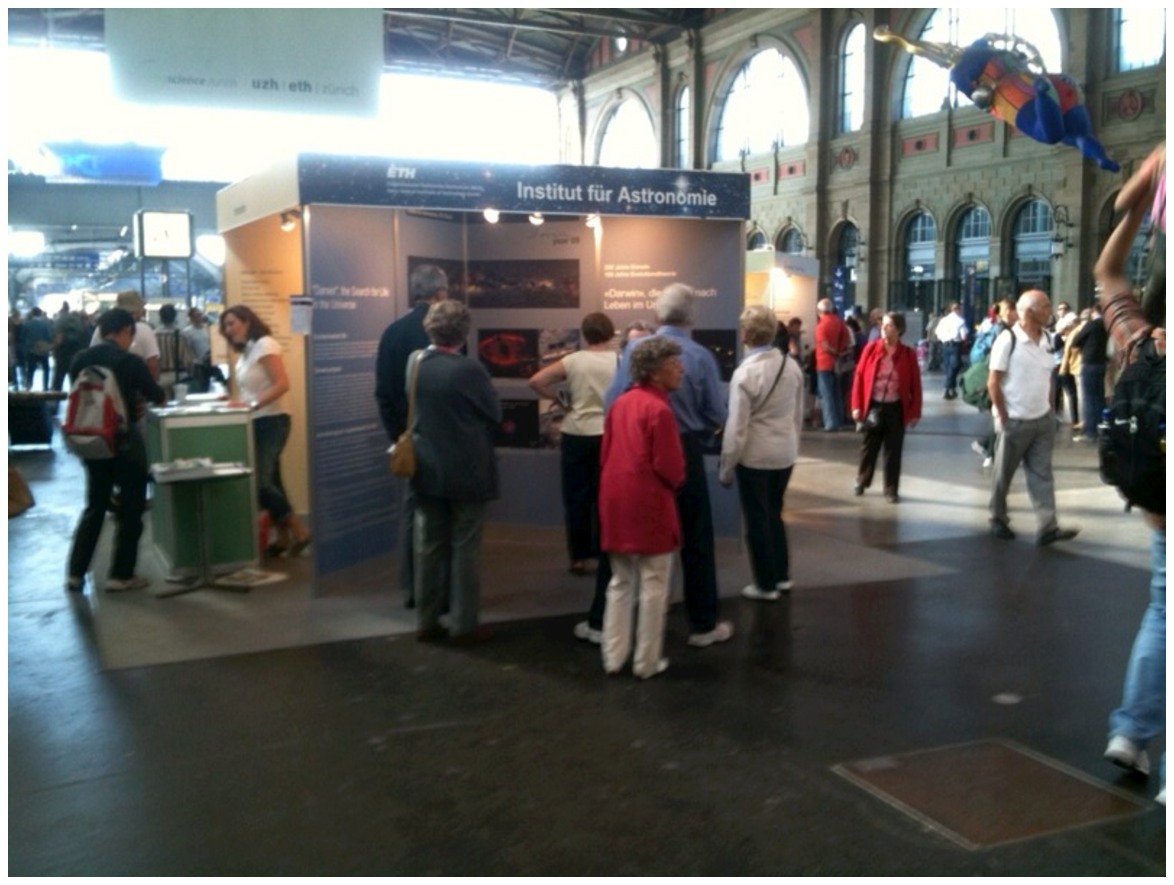}
\caption{An interested public in the Z\"urich Hauptbahnhof surveying ESA's plans to search for terrestrial planets around Sun-like stars.  The author fielded questions such as â ``how do you define life?â" and ``what sorts of planets are habitable?â" for which he had no good answers.  There was also a general admonition not to ``oversell" future missions to the public.}
\end{figure}

What are the initial conditions for planet formation?  These are constrained from observations that span the electromagnetic spectrum of primordial disks around young pre-main sequence stars in nearby star-forming regions.  Disk masses are inferred from (often) optically-thin millimeter wave emission to be between 0.001-0.1 M$_{\odot}$ (assuming standard gas-to-dust ratios).  Observed mass surface density profiles range from  $\Sigma \sim r^{-p}$ with $0.0 < p < 1.0$ (rather than p = 1.5  which is often assumed; see Andrews et al. 2009 and Isella et al. 2009).   It is generally accepted that disks exhibit inner holes, variable gas to dust ratios, as well as hydrostatic flaring (Pinte et al. 2008; Cortes et al. 2009).  Increasingly complicated models are needed to calculate in a self-consistent way their thermal and geometrical structure, not to mention dynamics(Dullemond et al. 2007).  These gas-rich disks exhibit large ranges of other properties as well including outer radii, accretion rates, and even lifetimes.  It remains to be demonstrated whether this diversity in initial conditions is responsible for the diversity observed in extra-solar planets (Meyer, 2009).

What should we expect from theory regarding the products of the planet formation process?  The basic ideas for collisional growth of planetesimals were outlined by Safronov in the late 1960s.  Refinements by Wetherill (1991) and many others are roughly consistent with observations (see below).  Planet formation should proceed more quickly around stars of higher mass due to both higher disk mass surface densities and shorter orbital timescales.  Yet disks around more massive stars appear to dissipate more quickly.  It is not yet clear which process wins out in planet formation.  

In what follows, I briefly review some successes but concentrate on failures in our understanding of planet formation in the hope of motivating the work needed to define near-term missions as well as realizing our ultimate aspirations within the Blue Dots initiative.

\section{Evolution of Gas Disks}   

All Sun--like stars probably form with some sort of circumstellar disk surrounding them (see for example Figure 2).  These disks are bins for storing angular momentum and provide the initial conditions for planet formation.  The `typical' lifetime of such disks is 1--3 Myr with large dispersion (Mamajek 2009).  In this section we discuss the evolution of solids, the formation of gas giant planets, aspects of disk chemistry, as well as
the final stages of disk dissipation. 

\begin{figure}
\plotone{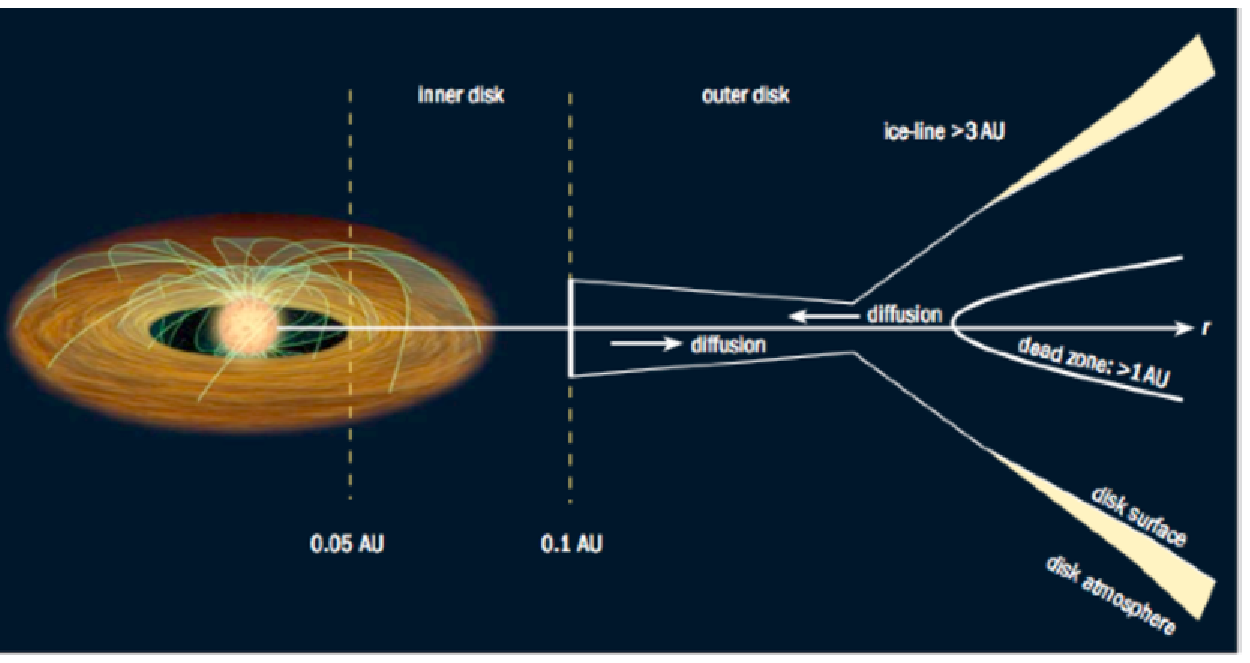}
\caption{Schematic of a circumstellar accretion disk adapted from Dullemond et al. (2001) incorporating artwork of 
R. Hurt (NASA/JPL) as it appeared in an article by M. Meyer in Physics World, November, 2009.}
\end{figure}
\subsection{Draining of Solids}   

Theorists who study planet formation create their own problems.  One of the current challenges is to understand how to preserve solid materials in gas-rich disks from which the cores of gas giants as well as terrestrial planets form.   A potentially serious loss mechanism is due to the gas drag suffered by bodies in Keplerian orbits embedded in a gas disk that is partly supported by thermal pressure.  Such bodies would orbit faster than the gas, suffer a head-wind, and spiral into the star.  For example, a meter-sized body at 1 AU suffering this effect would be lost in less than one century (e.g. Weidenschilling, 1977).  But as pressure gradients take, they can also give.  Others have investigated whether density enhancements or vortices expected in a turbulent accretion disk could serve to concentrate particles.  Johansen et al. (2007) suggest these effects could lead to gravitationally unstable regions capable of producing large bodies (~100 km) very quickly.  If bodies can grow faster than they migrate, they can reach a size where the effect is not important and are saved.  Further work is needed to explore observations that would demonstrate empirically the existence of the problem as well as the range of mechanisms which can solve it.   Perhaps thermal/chemical discontinuities in the disk (like the ice line) are those places where solids can collect and speed-up planet formation (e.g. Kretke \& Lin 2007). 

Another thorny issue is Type I migration where lunar mass and larger bodies launch spiral waves in a gas--rich disk.  Numerous torques permit the proto-planet to exchange angular momentum with the gas and migrate.  Calculations assuming a previously favored mass surface density profile $\Sigma(r) \sim r^{-3/2}$ led to the conclusion that the co-rotation torques are small, in agreement with linear theory.  In this case the inward migration rate is very rapid, leading to concerns about the viability of planet formation that depends on a healthy population of these embryos.  \citet{2005A&A...443.1067N} has explored whether scattering of migrating protoplanets off of density enhancements in the inner accretion disk could save these building blocks.   Recent calculations by Paardekooper \& Papaloizou (2009) have pointed out that co-rotation torques are non-zero for a range of relevant surface density profiles calling into question the ubiquity of rapid inward migration.  Models that produce planets favor migration rates an order of magnitude smaller than the early estimates (Ida \& Lin 2008).

Current models of Solar System formation start with a disk of gas and dust that cannot be much more than 10\% the mass of the young Sun.  With a standard gas to dust ratio that implies no more than a few hundred Earth-masses of material in heavy elements.  Because the current Solar System has $\sim$100 Earth-masses of solids, any methods of removing solids cannot have been extremely efficient. 

\subsection{Giant Planet Formation}   

Recent reviews of giant planet formation models can be found  in Durisen et al. (2007) and Lissauer \& Stevenson (2007).   To say that there is great controversy among experts regarding how giant planets form is an overstatement.  Two mechanisms proposed gather most of the attention primarily because the differences in the predictions appear to be stark in at least two important respects.  In the theory of gravitational fragmentation, no core (of several Earth-masses in heavy elements) is needed while in the core accretion theory, there is a core.  Further, gravitational fragmentation should operate primarily at large radii where the disk is cooler while core accretion would seem to favor formation at smaller radii (perhaps at the ice line).  The expected differences in planet composition are murky at best (Helled \& Schubert 2009).  Nonetheless, many researchers favor the core accretion theory, in part because of the expected correlation of gas giants with host system metallicity (Santos et al. 2001; Fischer \& Valenti 2005; Sozzetti et al. 2009).  Yet the planets we can most easily image directly are those at large radii (e.g. Marois et al. 2008, Kalas et al. 2008, and Lagrange et al. 2009).  It would appear likely that planets can form through gravitational instability though this is probably not the main pathway for gas giant planet formation (Boley, 2009; Rafikov, 2009).   And yet, they move.  The physics of Type II migration differs from Type I in that the proto--planet in question is large enough to open a gap in the disk.  The net effect is inward migration towards the star on the viscous timescale.  Evidence for this comes from multiplanet systems where one planet appears to have been caught in resonance with an outer more massive planet during migration (e.g. Kley et al. 2005). 

\subsection{Disk Chemistry}   

Sophisticated models for the temperature and density structure of circumstellar disks are starting points to investigate chemical processes within those disks and ultimately, the composition of planets that form within them (Woitke et al. 2009).  Heroic work in analyzing spectra from the Spitzer Space Telescope has revealed a number of molecular species including many emission features due to water vapor (Salyk et al. 2008; Carr \& Najita 2008).  Near-infrared echelle spectra obtained with 6-10 meter telescopes can provide velocity as well as spatial information (Pontoppidan et al. 2008).  Longer wavelength observations with Herschel and ground-based millimeter-wave telescopes trace cooler gas.  Combining these techniques could in principle yield molecular abundances as a function of disk radius.  Discontinuities in these abundance gradients could shed light on special radii in the disk where physical processes relevant to planet formation occur.  For example, compared to the Sun, the terrestrial planets in the Solar System are underabundant in carbon with respect to silicon by a factor of x 20.  If a significant fraction of the carbon from the interstellar medium (ISM) is delivered in the form of hydrogenated amorphous grains (HACs or even PAHs), then some process that transmutes this carbon to another form must be identified to explain why these grains do not contribute significantly to the bulk composition of the terrestrial planets.  Gail (2002) suggests that combustion reactions at temperatures above 800 K could decrease the solid carbon abundance substantially.  Lee et al. (2010) have provided an alternate explanation.  Whatever the solution, chemistry matters in determining the structure, geometry, and evolution of the disk and ultimately the composition of forming planets.  

Molecular abundances containing volatile species can also place constraints on the radial migration of icy solids into the inner disk (e.g. Ciesla \& Cuzzi 2006).   If Type I migration is responsible for transporting C--N--O rich icy planetesimals into regions where temperatures are high enough to liberate these species into the gas phase, we may see novel chemical abundances as a result.  In a pioneering study using "chemistry" as a probe of disk conditions, Pascucci et al. (2009) report differences in the C$_2$H$_2$/HCN emission line ratios between the disks surrounding solar-mass T Tauri stars and young brown dwarfs.  They speculate that the observations can be explained with a model where molecular nitrogen, photo-dissociated by UV radiation, drives HCN production more rapidly around higher mass stars with stronger accretion luminosity.  These studies are in their infancy but hold great promise for unlocking some secrets of planet formation.

\subsection{The Final Stages}   

There is a general concensus that gas-rich disks around Sun-like stars dissipate their inner disks on timescales of 1-10 Myr with a concomitant cessation of accretion onto the star (see Meyer 2009 and references therein).  Further there is growing evidence that gas-rich disks persist longer around stars of low mass (cf. Mamajek 2009), although there is some controversy concerning the duration of this transition phase.   Currie et al. (2009) argue for an extended transition phase of millions of years while Luhman et al. (2010) derive a transition phase of $< 10^6$ yrs consistent with several previous studies.  Estimates are made of the fraction of objects observed ``in transition" (from optically-thick to optically-thin) as well as the mean age of the sample.  The duration of the evolutionary phase is derived from the product.  Part of the discrepancy can be traced to the normalization of the fraction to the whole population versus the number of stars with detected disks\footnote{Thought experiments regarding the possible outcomes under a range of assumptions suggest that the former is preferred.}.  More substantive issues concern which types of objects are defined as ``in transition" and the fact that different regions sampled appear to yield different results.  Some studies find disks that are either: i) optically-thick at all wavelengths; ii) thick at long wavelengths and optically-thin at shorter wavelengths; or iii) thin at all wavelengths.  Other researchers find evidence for a continuum of disk spectral energy distributions with some exhibiting very modest excess emission over a range of wavelengths perhaps indicating flatter disk geometries (or ``homologously depleted'' disks of lower mass).  Self-consistent analysis of a number of data sets could resolve apparent contradictions.  

The issue is one of substance as models of disk dispersal make different predictions concerning the duration of this phase of evolution.  New observations with the Herschel Space Telescope of atomic fine-structure lines of [CII], [OI], and [NII] are sensitive probes of remnant gas at large radii.  UV and x-ray radiation ionizes and photo-evaporates a thin layer of exposed material on the disk surface (Gorti \& Hollenbach 2008; Ercolano et al. 2009).  PACS observations should be sensitive to fractions of an Earth mass in remnant disk gas around nearby young stars (e.g. Augereau et al. 2008).  The ability to observe the "last gasps" of these disks as they dissipate could be crucial to understand the formation of the ice giants Uranus and Neptune in our own Solar System.

\section{Rocky Bodies:  Fire and Ice}    

As the gas goes away we turn our attention to remnant dust created in collisions of larger parent bodies.  Indeed T Tauri disks themselves may already bear witness to collisions of kilometer-sized planetesimals as the dust that dominates the opacity we see may be second-generation rather than pristine ISM grains.  Nonetheless, in the gas-poor debris phase, observing this dust helps us to trace the evolution of planetesimal belts which in turn build rocky planets from the material left over after the end of the gas-rich accretion disk phase (e.g. Meyer 2009 and references therein).

\subsection{Tracing the Formation of Terrestrial Planets}   

It is certainly tempting to link dust generated in collisions in debris disks with processes responsible for planet formation.  An observational starting point is to determine the frequency of infrared excess emission as a function of stellar age and compare the results to recent models of planet formation (e.g. Siegler et al. 2007; Currie et al. 2008).  Fortunately for observers, some modeling groups take the extra step of estimating this expected infrared excess as a function of input assumptions facilitating comparisons to 
the data (e.g. Kenyon \& Bromley 2008).  Yet a flux measured at a single wavelength can be explained in many ways.  Meyer et al. (2008) interpreted the evolution of 24 micron excess emission observed towards Sun-like stars in terms of a warm dust model related to the formation of terrestrial planets.  Detectable mid-IR emission decreases substantially around Sun-like stars at ages $>$ 300 Myr.  However, it is not clear whether this evolution is in the dust temperature or in the overall dust luminosity.  Carpenter et al. (2009) were able to test this using spectro-photometry from 5-30 microns and found that the dust was cooler than typically assumed.  This implies planetesimal belts located beyond 3 AU with a mode in the distribution of inner radii of 10 AU.  There is no evidence for temperature evolution and modest evidence for luminosity evolution of the dust for ages 3-300 Myr.

Occasionally one detects objects whose IR excess is so large, it cannot be explained with any reasonable quasi-equilibrium debris disk model (Wyatt et al. 2007).  In these cases one invokes the uncomfortable hypothesis that the phenomena is short lived.  Because we observed only the product of the duration and frequency of the events, it is difficult to interpret the statistics in the context of planet formation through giant impacts (Nagasawa et al. 2007).   While the frequency of detectable mid-IR excess emission around Sun-like stars is about 10-20\% for ages between 3-300 Myr, fewer than 10\% of these exhibit evidence for transient emission.  If we hypothesize that every sun-like star has 1-3 terrestrial planets, and each is built through a series of 3-10 giant collisions, we should expect to see dozens of collisions during the epoch of terrestrial planet formation (10-100 Myr).  If each collision produces a cascade of debris that persists for $10^{5}$ years, the observed statistic of 1-3\% can be explained.  Note that a duration of $10^{5}$ years is much longer than the expected lifetime of small dust grains (Wyatt, 2008), so an extended phase of dust generation would be required. 

What are the expected observational signatures of such giant collisions?  Some transient systems exhibit unusual spectral features due to amorphous silica thought to be produced only in high velocity impacts (e.g. Rhee et al. 2008)\footnote{Silica is also observed toward a handful of T Tauri stars (Sargent et al. 2009).}.  Lisse et al. (2009) suggest that the Spitzer spectra of HD 172555 contains evidence for both silica and SiO gas in the system.  Such emission could be produced by a giant impact in which some fraction of the impactor was vaporized.  In the simulations of Canup (2004) about 5$\times$10$^{-3}$ Earth-masses of hot gas are released in the impact of a Mars-sized body with the proto-Earth.  If the vapor condensed into micron sized grains, this would result in mid-IR emission 100$\times$ above the Spitzer detection limit.  However such dust might only persist for 10$^3$ years (around stars of age $10^{7}$) before removal by radiation pressure.  Unless subsequent collisions prolong the dust generation phase (not to mention the expected duration of gas emission), such objects are expected to be very rare.  SiO gas emission has only been claimed for a single object  (and awaits ground-based confirmation).

Could we detect directly a forming terrestrial planet, the result of such a giant impact?  Zahnle et al. (2007) review the geophysical consequences of the giant impact hypothesis for the evolution of the early Earth.  Pahlevan \& Stevenson (2007) predict the existence of a large magma disk ($>$1500 K!) of a few Earth radii that would persist for thousands of years (Figure 3).  An object of such high temperature and large solid angle would be easily detectable (if resolved from the host star).  Finally we note that Miller-Ricci et al. (2009) have explored whether we could directly detect molten Earths or super-Earths.  Current technology would enable us to detect such bodies beyond 30 AU around stars in typical search samples, but they appear to be rare.  New instruments that employ novel diffraction suppression strategies on existing 6-10 meter telescopes (e.g. Kenworthy et al. 2007) as well as the next generation ELTs present exciting possibilities.

\begin{figure}
\plotone{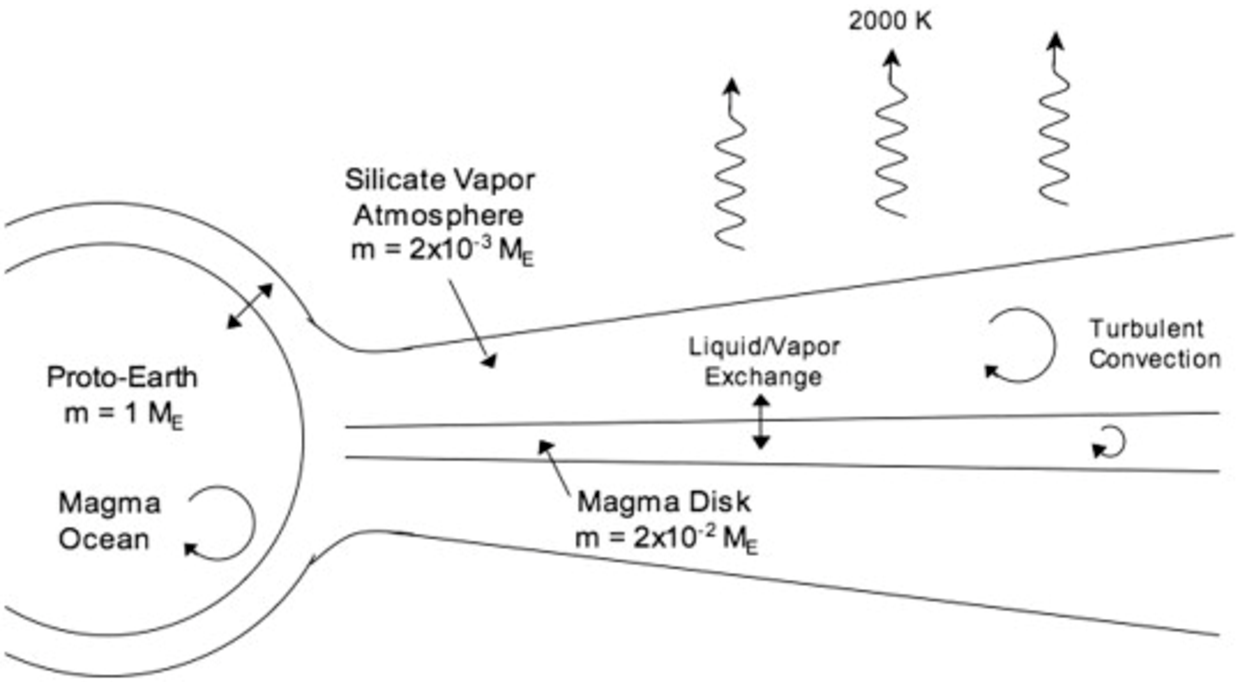}
\caption{Schematic of the magna disk surrounding the proto-Earth after the Moon-forming impact.  Such a large hot disk would be easy to detect around near-by stars (from Pahlevan \& Stevenson 2007)}
\end{figure}

\subsection{Dynamics = Composition}   

Physical models of circumstellar disks inform calculations of disk chemistry telling us what can be where at a given time.  Fluid dynamics governing the motions of solids in gas disks can move things around.  But so can gravity alone after the dispersal of the gas.  Raymond et al. (2004) assessed what fraction of volatiles (such as water) could end up as part of forming terrestrial planets built through collisions of planetesimals that originate from a variety of locations.  The basic idea is that static temperature considerations (such as the location of the ice-line) can tell you where water-rich planetesimals might be located in a disk.  And the dynamics of N-body systems can help you assess where they end up.  A general feature of N-body calculations of collisions capable of forming terrestrial planets within a few AU of Sun-like stars is that their eccentricities and inclinations are generally too high.  O'Brien et al. (2006) have explored dynamical friction from swarms of small bodies to dampen the eccentricities with some success (see also Morishima et al. 2008).

Bond et al. (2010) have taken this approach one step further combining the results of a time-dependent accretion disk model with a chemical condensation sequence model to predict the composition of planetesimals as a function of radius in the disk (assuming no bulk migration of solids).  Convolving the output of these calculations with a dynamical code that tracks what fraction of each planet formed from the initial annulus in the disk, one can estimate the bulk elemental abundances of the planets.  The approach, though simplistic, does a remarkably good job for the abundances of many elements in the Earth and Venus (excluding volatiles).  Future work exploring similar models as a function of stellar mass, stellar composition, comparing results from other codes, and including the effects of solid migration could provide additional insight. 
A potentially observable test of such models could be the amount of heavy elements deposited in the stellar photosphere after the radiative core of the Sun-like star develops.  Models can predict the amount of solid material lost into the star from collisions of rocky planetesimals.  These calculations can be compared to the dispersion  in heavy element abundances observed in young clusters such as the Pleiades.  Observations reported by Wilden et al. (2002) indicate that fewer than 5 Earth-masses of refractory elements could have been accreted in a differential sense.  This limit approaches the predictions from models and future work could provide more stringent constraints.  Finally we note that, contrary to pollution scenarios, Melendez 
et al. (2009) find evidence that the Sun is depleted in refractory elements compared to other Sun--like stars, perhaps
offering another tool to pre--select ``systems like our own'' for closer scrutiny.

\subsection{Evolution of Outer Belts}   

Wyatt (2008) and references therein recently review the evolution of debris disks around stars as a function of stellar type.  In general one can say that debris disks (of higher mass) appear more common around A stars than around G stars or M dwarfs (Su et al. 2006; Trilling et al. 2008; Carpenter et al. 2009; and Gautier et al. 2007).  There is some evidence for luminosity evolution in that older stars sport fainter debris though there is a large dispersion in most observed quantities.  One striking feature of the data reported for Sun-like stars from the FEPS project (Carpenter et al. 2009) is that inner disk hole sizes are typically 10 AU, even for stars 3-30 Myr old.  Morales et al. (2009) also report large inner holes for A star samples.  Whatever clears these debris disks out beyond 10 AU (perhaps efficient planet formation) happens relatively fast.  A healthy fraction of these debris disks cannot be fitted with single temperature models implying debris over a range of radii.  It remains to be seen whether models of extended debris (like a modified T Tauri disk model; Hillenbrand et al. 2008) fit as well, worse, or better than models with two distinct inner and outer belts (like scaled versions of the asteroid and Kuiper belts; Su et al. 2009).  

It is clear that Spitzer observations have detected only the tip of the iceberg regarding dusty debris around Sun-like stars.  Bryden et al. (2006) estimate that the data are consistent with the median of Sun-like stars having a debris signature within $\times$ 10 of our own (see also Greaves et al. 2010).  Recent models for the evolution our own Solar System evolution invoke migration of the giant planets as a key feature about 900 Myr after formation (Morbidelli et al. 2009).  Jupiter and Saturn migrate inward, and Uranus and Neptune migrate outward, wreaking havoc on both the asteroid and Kuiper belts in the process.  This dynamical depletion of dust-producing parent bodies is perhaps the main reason why our system is so faint today (Booth et al., 2009; Meyer et al. 2007).  Other planetary system architectures might be expected to undergo comparable dynamical rearrangements at times determined by their configurations.

Is there a connection between outer debris disks traced by far-IR/mm-wave emission and extra-solar planets detected via other means?  Moro-Martin et al. (2007) searched for a correlation based in disk samples from the Spitzer Space Telescope and found none (see also Bryden et al. 2009).  Yet there are some very notable exceptions:  HD 69830, eps Eri, Beta Pic, Fomalhaut, and HR 8799.  Su et al. (2009) point out the unique spectral energy distribution of HR 8799 may hint at disk sculpting by planets.  Imaging such disks in scattered light (e.g. SPHERE and GPI) , and in thermal emission (Herschel, ALMA, JWST) will provide fundamental constraints on models needed to make accurate estimates of their physical properties.

\subsection{Are Planetary Systems Common?}   

Radial velocity surveys suggest that 8.5\% of Sun-like stars possess planets with masses $> 0.3 M_{JUP}$ within 3 AU (Cumming et al. 2008).  Extrapolations of those results beyond about 30 AU are inconsistent with existing null results (e.g. Nielsen \& Close, 2009).   The next generation of imaging surveys will test the prediction that $\sim$ 20\% of Sun-like stars have gas giants inside of 20 AU (e.g. Cumming et al. 2008).  There are already hints from on-going RV work as well as micro-lensing statistics that lower mass (3-10 Earth-mass) planets are even more common (Gould et al. 2010; Mayor et al. 2009).  
If we take these results as confirmation of the planetary population synthesis models, we can speculate that terrestrial planets will be more common still (Mordasini et al. 2009; Ida \& Lin 2004).  
Finally we note the recent work of Wright et al. (2009) who point out that at least 28 \% of known planetary systems contain multiple planets (like our own).

And if terrestrial planets are so common, can we detect them?  Above, we explored a high-risk scenario for detecting hot terrestrial planets during their collisional formation.  Absil et al. (this volume) present observations of reflected light from inner zodiacal dust emission which could present challenges for the direct detection of terrestrial planets.  The Herschel Space Telescope as well as the LBT Interferometer will be powerful tools to explore cool as well as warm debris down to levels $\times$ 10 that inferred for our planetary systems (Hinz 2010).  The case for a 1-2 meter space-based visible coronagraph capable of detecting debris down to Solar System levels seems clear.  Such a facility could also detect Earth-like planets around a handful of nearby stars as well as characterize dozens to hundreds of super-Earths and gas giants in visible reflected light (e.g. Guyon et al. this volume).  The joint JAXA/ESA mission SPICA could also play a crucial role in planning for a future space-based mission (Goicoechea et al. this volume).  Finally, we note that the next generation of extremely large telescopes might also take the first image of a terrestrial planet around a nearby star (Kasper et al. 2009). 

\section{Implications}   

One thing is very clear from studies concerning the formation and evolution of planetary systems:  all disk properties exhibit great dispersion at a given time.  Yet some key parameters appear to be a function of star mass (like initial disk mass and gas-disk lifetime).  While the draining of solids represents serious challenges to current theories of planet formation, these physical mechanisms may also present opportunities for planetesimal growth.  Regarding the formation of giant planets, some lines of evidence favor the core accretion theory, yet it seems clear that gravitational instability might also play a role.  It may be that hybrid approaches are needed where gravitational instabilities collect solid material in gas rich disks enhancing the prospects for the core accretion scenario.  Available observational constraints (meager though they are) are consistent with terrestrial planets being very common.  We may even be lucky to catch some in the process of collisional formation! 

Many open questions remain.  Does planet formation favor dynamically dense systems?  Perhaps all planetary systems are packed in the sense of being stable only on timescales comparable to their current age (Laskar 1994; Barnes \& Greenberg 2007).  If so, it is those systems lacking circumstellar debris that could harbor the richest planetary architectures.  Is planet formation in multiple star systems different than around single stars?  We are only beginning to tackle this question from theory (Quintana et al. 2007) and observations (Eggenberger 2009).  How does planet formation vary from one star forming environment to another?  Studies of disk evolution as a function of cluster environment are on-going (e.g. Mann \& Williams, 2009).  The presence of massive stars (likely in a rich cluster) could have important effects on the planet-forming potential of disks found around nearby sun-like stars (Adams et al. 2006).

If we could perform an on-line search to get directions for a ``Pathway Toward Habitable Planets'' the results might read something like this: 1) Straight AHEAD: with medium class missions capable of answering key questions in the next decade; 2) 'RIGHT': strong exoplanet science cases for a variety of ELT instrumentation; 3) Don't be LEFT behind:  invest now in diverse technologies for a flagship mission in order to save money when final design concepts are needed; 4) WATCH for:  differences between exploration and scientific hypothesis testing in developing 'level one' requirements.  Will you learn something from the null result?  Such an investment should teach us something fundamental about the formation and evolution of planetary systems; and 5) Don't STOP:  until you reach your goal of imaging and taking spectra of terrestrial planets around nearby Sun-like stars!

\acknowledgements 
I would like to thank the organizers of the Blue Dots initiative for staging such a stimulating conference in an inspiring setting as well as colleagues on the Working Group for the Formation and Evolution of Planetary Sytems for sharing their ideas.  Special thanks to I. Ribas for his patience and advice in preparing this manuscript and H. Zinnecker for providing comments on a draft version.

\end{document}